# Parallel Data Compression Techniques

David Noel, Elizabeth Graham, and Liyuan Liu
The University of Florida ECE Department
Gainesville, FL, USA

**Abstract**

*With endless amounts of data and very limited bandwidth, fast data compression is one solution for the growing data-sharing problem. Compression helps lower transfer times and save memory, but if the compression takes too long, this no longer seems viable. Multi-core processors enable parallel data compression; however, parallelizing the algorithms is anything but straightforward since compression is inherently serial. This paper explores techniques to parallelize three compression schemes: Huffman coding, LZSS, and MP3 coding.*

## I. INTRODUCTION

Data compression is not supposed to be done in parallel. This is because the order is very important to file viability and accuracy. Textbooks could be compressed more by making the font microscope or moving words around so there is less wasted space, but obviously, a book that is impossible to read (both visually and conceptually) is useless.

There is a limited amount of digital and actual space available to store everything, but more importantly, larger files can have exceptionally long transfer times. The simple solution is to make them smaller. Generally, the higher the compression ratio, the longer it takes to compress and decompress it. So you can spend a substantial amount of time waiting for the file to compress/decompress and a relatively shorter transfer time or wait for it to upload/download.

For this project, we researched how to improve the speed and compression ratios of three popular serial compression techniques by implementing them in parallel: Huffman coding for lossless fixed-sized alphabets (e.g., text files), LZSS for lossless variable-sized alphabets (e.g., bitstreams), and MP3 coding for lossy coding (e.g., digital music).

## II. RELATED WORKS

Most recent data compression algorithms utilize some form of lossy-compression scheme [1], i.e., the original data cannot be completely reconstructed from the algorithms. While this method may be very fast, producing very favorable compression ratios, it is not acceptable for some situations where the information to be compressed is critical. Such information exists in text, medical, and executable files, whose original meaning and contents must be maintained after compression.

This part of the project will focus solely on the parallelization of compressing text data. The form of compression algorithm utilized to achieve such tasks is lossless, i.e., they reduce the storage requirements of the files or time required to transfer the files while maintaining the original meaning of the file contents.

Lossless compression uses statistical redundancy inherent in most text files or files whose contents represent real-world data. Examples include biological data, where certain terms are consistently repeated, or MRI images, where large chunks of the image are colored "black." The most widely used form of lossless algorithms is the Lempel-Ziv (LZ) compression algorithms [4][2], and the associated tables or dictionaries they create are usually Huffman encoded.

Since the inception of the LZ methods, several variants have been developed [1]; all aimed at more efficient compression ratios. Some of these new variants include LZW (used in GIF images), LZR (used in Zip files), and LZX (used in CAB files). However, some of the most current algorithms use probability predictions, grammar-based codes, or combining statistical predictions with arithmetic coding methods [4].

In terms of audio compression, we want to compress an original uncompressed audio file using an MP3 encoding scheme. An audio file records what the audio waveform is over time. If the sampling rate is 44.1 kHz, there are 44100 points, in a type of double or float or integer depending on the file format, a second, which is uniformly distributed, to describe the waveform. [6][8][10] So the basic type of audio file is filled with numbers. PCM, or Pulse Code Modulation, is a standard format for uncompressed digital audio CDs. [9] The WAVE file format is a common container for PCM data. [10]

MP3, or MPEG-I Layer III, is an audio compression standard. It exploits psychoacoustics facts, such as that most people usually cannot hear audio outside of the 20 Hz~20 kHz band (or even can't further than 16 kHz), and Huffman coding to do audio compression. [12] Other techniques, such as joint stereo effect, can be optionally applied to reduce the resulting file size or quality.

Some literature describes parallel MP3 compression using several DSPs: Some slave DSPs complete sub-band analysis and psychoacoustic modeling, while one master

DSP executes bit allocation, quantization, and bitstream formatting[6]. This could also be done on multi-threaded, multi-core, or connected machines. Another approach exploits data parallelism.

### III. APPROACH

For each data compression scheme, two methods for parallelization were explored along with the serialized version.

*A. Huffman Coding*

This approach to data compression was primarily focused on *text* data, where the dataset utilized was a 45MB text file that was a combination of the complete text versions of *The King James Bible*, the *2012 Encyclopedia Britannica*, *A Tale of Two Cities* by Charles Dickens, and James Joyce's *Ulysses*. The Algorithms developed were implemented on both Iota and Beta clusters, and their relative performance was compared and discussed.

METHOD 1: This method implements a Sequential Huffman Coding scheme, a fairly efficient method for compressing text data as it takes $O(n \log n)$ time. It is a lossless, variable length-encoding method of compression where a small number of bits are used to encode characters based on their probability of occurrence. The results are then processed using an extended weighted binary tree, replacing two external nodes with one internal node. The weight of this internal node is the sum of the weights of the two external nodes. The procedure is then repeated until the root node is reached.

METHOD 2: In this method, we will seek to parallelize the sequential Huffman Coding developed in Method 1. Our approach would be first to segment our dataset into discrete segments, each with a unique id proportional to the number of processes used. Each segment would then be sent to a different process where the Huffman algorithm would be implemented by that process in order to compress the segment it received. Upon completion, each process would then send its result to the parent process, where the segments will be re-assembled based on segment id and stored in a separate file as encoded compressed data.

METHOD 3: In this method, we modified METHOD 2's Parallelization scheme by first running the Burrows-Wheeler Transform (BWT) on our data set– this transformation reverses the order of a block of text using well-established sorting algorithms such as Quicksort. Upon transformation, the dataset would consist of several places where a single character is repeated multiple times. This then lends itself appropriately to huge gains in compression and performance parameters since it would be much easier to compress a string that has runs of repeated characters. After the said transformation was done, we applied the Huffman Coding algorithm developed in Method 1 to the result to compress the data in parallel.

*B. LZSS*

The Lempel-Ziv-Storer-Szymanski (LZSS) compression algorithm is a popular method for compressing data of any length (where Huffman coding is limited to a finite alphabet). LZSS saves space compared to other approaches because it does not store a dictionary. Instead, it is based on literals and references (address and length) to the file itself.

Compression begins with a known window (e.g., all zeros, counting numbers) of some user-specified length. Starting with the first bits of the file, the algorithm searches for a match of at least some minimum length in this window. If one is found, a 1 is written, followed by the address offset of the first matched bit in the window and the length. The number of bits allotted for the offset and length are fixed to allow for decompression. If no match is found, a 0 is written, followed by the literal value. Referenced strings are added to the end of the window, and older strings are rotated out.

Unfortunately, since the main idea behind the algorithm is searching, it is a slow compression scheme. Additionally, since the window and the string to be matched completely depend on the results of compression earlier in the file, parallelizing the search is not simple.

METHOD 1: The serial version of standard LZSS, plus a few additions found to make LZSS better. First, a user-defined block size was added. It is possible for a file with little or no repetition actually to double in size with this algorithm. To help minimize the effect, block size indicates how many bits are included in a literal. The higher the block size, the fewer the necessary encode bits. Second, the ability for the file to specify the initial window was included. The first bytes of a file are far more likely to predict future data in the file than a static one. The first X bits (enough to fill the window) in the file remain as they are (un-compressed) to be used for decompression.

METHOD 2: This is a segmented parallel version. The file is divided into equal chunks for each processor, and each processor independently compresses its chunk. Using a pre-defined number of bits to define the length of each compressed chunk (since they are most likely to be different sizes afterward), pack the lengths and compressed data together in order.

METHOD 3: This is a cooperative dictionary construction (CDC) parallel version. Each processor has a chunk of the window (the chunks overlap to prevent processors from communicating too much with each other) within it searches for the not-yet-coded string. The master processor collects the best results from the remaining processors, determines the longest match, and encodes the data accordingly.

*C. MP3 Coding*

There is one serial version and two approaches for MP3 coding: segmented parallelization and work-sharing parallelization.

METHOD 1: In the serial version, the compression will go through the standard MP3 compression process one by one: Firstly, do psychological modeling to strip out unused frequency bands (over 17 kHz in our case); secondly, do MDCT to convert samples to easily quantized sets of 576 frequency domain samples; thirdly make quantization loops to quantize converted frequency domain samples and do Huffman encoding, and make sure the target bit rate is achieved; lastly combine all the audio data, generate header contents and CRC codes to do frame forming.

METHOD 2: In the segmented parallelization version, all worker processors are equal, running the same code, except for the master processor, which is always processor 0. The master processor will first read the header of the source audio file, and determine some basic parameters, such as the file size, the number of audio frames, and the number of processors. Then it will decide how to scatter the source audio file to other processors. Each processor will do MP3 compression independently. Each channel will be encoded separately. Then the primary process will gather all the parts and produce a single MP3 file.

METHOD 3: Another approach is called work-sharing parallelization. In an MP3 file, audio data are stored in many *frames*. Each frame contains the same number of encoded audio samples. Moreover, there is stuff besides actual audio data in each frame, including the Huffman code table. This all means that each frame is independent of others; thus, we can encode them independently. In addition, two channels can be encoded separately. So the second approach is that: every two cores will encode one frame of audio data, which is 1152 samples, and each one of them will encode one channel, either the left channel or the right channel. If two more cores exist, one more frame will be encoded simultaneously.

## IV. EXPERIMENTS AND RESULTS

### A. Huffman Coding

The initial approach to this part of the project was the development of the best sequential algorithm possible, which would naturally lend itself to parallelization and achieve a compression ratio of less than 55%. After approximately 90 hours of non-consecutive algorithm and code development, an acceptable sequential algorithm and its parallel counterparts were produced that achieved the criteria above.

In order to obtain well-rounded results, each of the two parallel versions of the algorithms described above was implemented using MPI, MPI+OpenMP, and UPC for a total of 7 algorithms, including the sequential one.

The analysis began by using the Iota cluster to run the sequential program to completion and then recording its wall clock time (Latency) and amount of compression achieved. The first parallel version, described in METHOD 2 above, was then executed at the Node level in contrast to the core level. This allowed for the quick and efficient testing of multiple cores on multiple chips simultaneously while simplifying reporting purposes. The nodes were chosen in steps of $2^n$, where n=0, 1, 2, and since each node consists of 4 cores, the parallel algorithm was executed on a maximum of 16 cores concurrently (4 Nodes). The elapsed time and amount of compression were recorded at the end of each job execution. This process was then repeated for METHOD 3.

Similarly, the entire process described above was implemented on the Beta cluster, and results were recorded. The following tables and charts present the Huffman Coding and parallelization findings.

TABLE I. COMPRESSION LATENCY PER ALGORITHM ON BETA CLUSTER

| Algorithm | Latency (Seconds) | | |
|---|---|---|---|
| | *1 Node* | *2 Nodes* | *4 Nodes* |
| Method 1(Sequential) | 7.4105 | | |
| Method 2(MPI Only) | 7.879 | 8.349 | 4.855 |
| Method 2(MPI+OpenMP) | 7.8487 | 8.287 | 4.754 |
| Method 2(UPC) | 7.976 | 8.357 | 4.812 |
| Method 3(MPI Only) | 7.893 | 2.375 | 1.219 |
| Method 3(MPI+OpenMP) | 7.6745 | 1.939 | 1.308 |
| Method 3(UPC) | 7.949 | 2.488 | 1.244 |

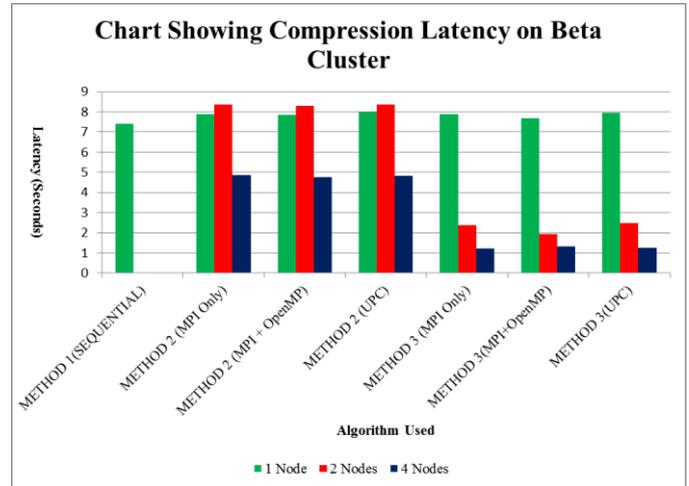

Fig. 1. Chart showing compression Latency on Beta Cluster

TABLE II. COMPRESSION LATENCY PER ALGORITHM ON IOTA

| Algorithm | Latency (Seconds) | | |
|---|---|---|---|
| | *1 Node* | *2 Nodes* | *4 Nodes* |
| Method 1(Sequential) | 5.5988 | | |
| Method 2(MPI Only) | 6.1449 | 8.691 | 5.086 |
| Method 2(MPI+OpenMP) | 5.9834 | 8.368 | 4.679 |
| Method 2(UPC) | 6.0624 | 8.526 | 4.671 |
| Method 3(MPI Only) | 6.189 | 0.732 | 1.261 |
| Method 3(MPI+OpenMP) | 6.34 | 2.316 | 2.758 |
| Method 3(UPC) | 5.723 | 2.39 | 1.283 |

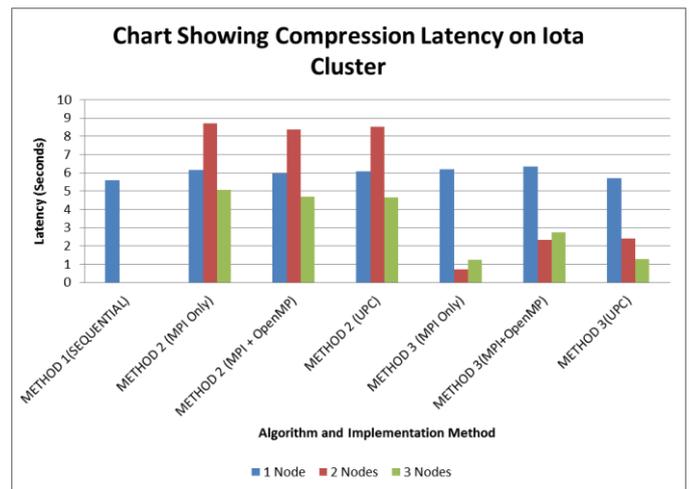

Fig. 2. Chart showing compression Latency on Iota cluster

TABLE III. COMPRESSION ACHIEVED ON BETA CLUSTER

| Algorithm | Compression (%) | | |
|---|---|---|---|
| | *1 Node* | *2 Nodes* | *4 Nodes* |
| Method 1(Sequential) | 48.26 | | |

| Algorithm | Compression (%) | | |
|---|---|---|---|
| | *1 Node* | *2 Nodes* | *4 Nodes* |
| Method 2(MPI Only) | 49 | 49.74 | 49.6 |
| Method 2(MPI+OpenMP) | 50.435 | 50.87 | 51 |
| Method 2(UPC) | 51.576 | 49.59 | 49.94 |
| Method 3(MPI Only) | 97.67 | 98.79 | 96.6 |
| Method 3(MPI+OpenMP) | 95.56% | 98.79 | 96.7 |
| Method 3(UPC) | 93.34 | 98.7 | 96.6 |

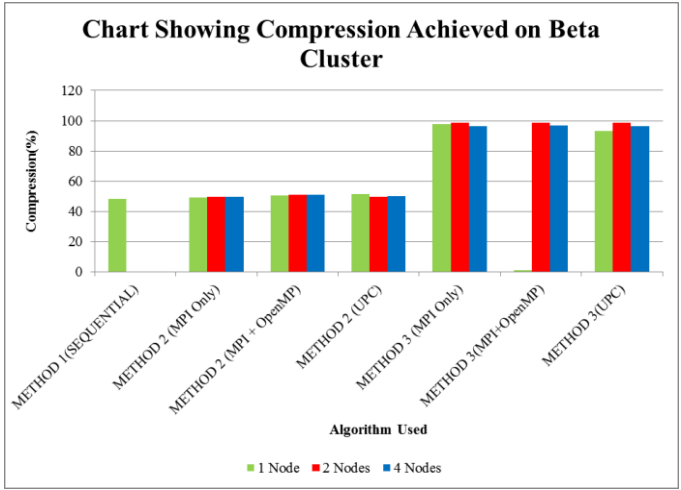

Fig. 3. Chart showing compression achieved on Beta

TABLE IV. COMPRESSION ACHIEVED ON IOTA CLUSTER

| Algorithm | Compression (%) | | |
|---|---|---|---|
| | *1 Node* | *2 Nodes* | *4 Nodes* |
| Method 1(Sequential) | 50.1 | | |
| Method 2(MPI Only) | 49.67 | 50.44 | 49.61 |
| Method 2(MPI+OpenMP) | 48.24 | 49.79 | 50.55 |
| Method 2(UPC) | 48.76 | 48.85 | 49.89 |
| Method 3(MPI Only) | 97.65 | 98.78 | 96.65 |
| Method 3(MPI+OpenMP) | 97.54 | 98.72 | 96.48 |
| Method 3(UPC) | 97.77 | 98.78 | 96.47 |

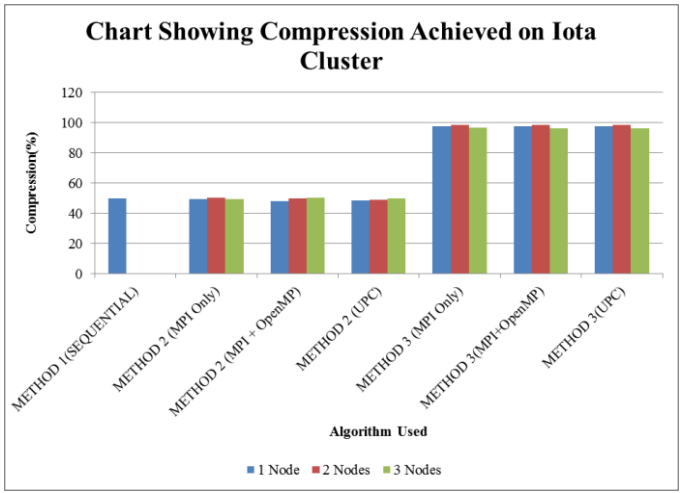

Fig. 4. Chart showing compression achieved on Iota

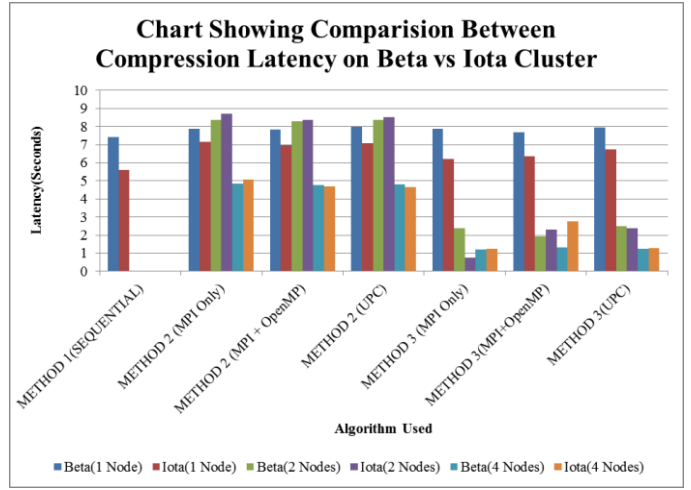

Fig. 5. Chart comparing latencies on Beta and Iota

*B. LZSS*

Methods 2 (LZSS Segmented) and 3 (LZSS CDC) exploit data-level parallelism differently. Both were implemented in MPI and tested, along with the serial version, with different LZSS attributes (window size, maximum match length, and minimum match length), initial windows, and the number of cores. Also, the effectiveness of each method will be juxtaposed with the time and complexity of developing it. Moreover, finally, each method will be tested on different types of files: Randomly-generated sequences, images, and FPGA bitstreams.

TABLE V. COMPRESSION RATIOS ON BETA

| Algorithm | Compression (%) | | |
|---|---|---|---|
| 1024/24/12/8 | *Random* | *Adder* | *Processor* |
| Method 1 with Dict 1 | -11.3 | 29.1 | 15.3 |
| Method 1 with Dict 2 | -7.7 | 43.4 | 27.5 |
| Method 2 with Dict 2 | -9.6 | 43.4 | 29.1 |
| Method 3 with Dict 2 | -8.3 | 89.75 | 58.4 |

Table V shows the compressions times from experiments run on 3 files (1 random sequence and 2 FPGA bitstreams). The notation "1024/24/12/8" at the top of the table indicates that the window size was 1024, the maximum match length was 24 bits, the minimum match length was 12, and the block size was 8. As for the windows, the initial values were either Dictionary 1, which is counting numbers, or Dictionary 2, which is the first 1024 (window size) bits of the file. The parallel measurements given were run on 4 cores.

TABLE VI. EXECUTION TIMES ON BETA

| Algorithm | Execution Time (Seconds) | | |
|---|---|---|---|
| 1024/24/12/8 | *Random* | *Adder* | *Processor* |
| Method 1 with Dict 1 | 0.017 | 18.3 | 42.9 |
| Method 1 with Dict 2 | 0.018 | 21.9 | 46.1 |
| Method 2 with Dict 2 | 0.017 | 5.8 | 12.11 |
| Method 3 with Dict 2 | 0.022 | 8.76 | 48.3 |

Table VI shows the execution times for the same experiment as Table V. It should be noted that the random sequence was 1000 B, the adder bitstream was 763 kB, and the soft-core processor bitstream was 1.8 MB.

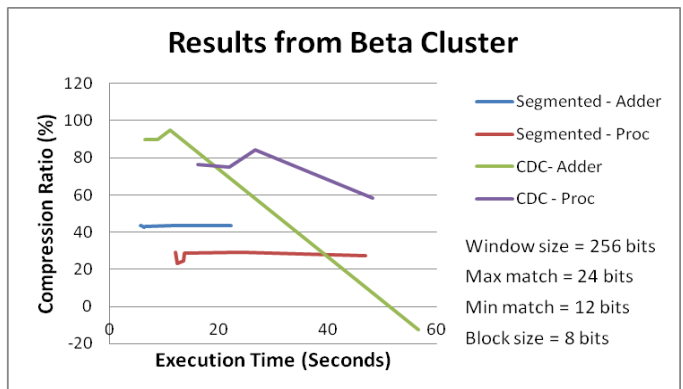

Fig. 6. Compares execution time vs. compression ratio for the adder and processor bitstreams.

TABLE VII. LZSS METHOD COMPLEXITY

| Method | Time to Develop Hours: Min | Lines of Code | Versions |
|---|---|---|---|
| Method 1 (LZSS Serial) | 30:47 | 367 | 2 |
| Method 2 (LZSS Segmented) | 22:12 | 549 | 3 |
| Method 3 (LZSS CDC) | 11:58 | 515 | 1 |

The time spent developing each specific method was recorded to examine the benefits of the parallelized versions further. Also worth noting, aside from Method 1, each method was partially completed using portions from earlier methods.

*C. MP3 Coding*

The audio file used as a source is a 120-minute WAVE file. The file is 1.2GB in size. Detailed parameters are shown in Table III. All experiments are done on the Iota cluster. For experiments 1) and 2), a program called LAME (version 3.99.5) is used. The LAME is open-source and is compiled on the Iota Cluster. For the rest of the experiments, we adapt the source code of LAME according to different architectures and methods and get it compiled.

TABLE VIII. SOURCE AUDIO FILE PARAMETERS

| Length | File size | Bit depth | # of channels | Bit rate | Sampling rate |
|---|---|---|---|---|---|
| 120 min | 1227 MB | Signed 16-bit | 2 | 1411 kbps | 44.1 kbps |

*1) Data Transmission Experiment*

A test program is developed to get a better idea of the system and the encoding process. Data transmission rate from the disk where the audio file resides to the computing nodes is calculated, including reading continuously and reading many cycles, with each cycle a few bytes to be read. The encoding time for one audio frame is also calculated.

The read speed from the disk to the computing nodes is shown in Fig.6.

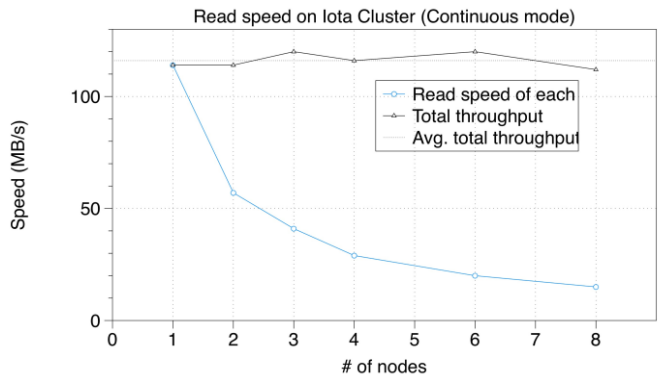

Fig.7. (a) Read speed in normal continuous mode.

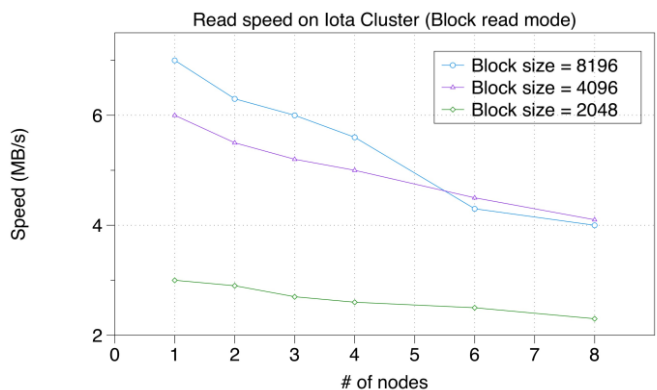

Fig. 7. (b) Read speed in block read mode.
Fig.7. Read speed in two modes.

From Fig.7, we can get a few observations: 1) Data reading in normal continuous mode is much faster than in block read mode. 2) The bigger the block is, the faster the speed is. 3) In continuous read mode, the speed of each node declines almost linearly as the number of nodes increases. However, the total throughput stays the same. This is likely to be limited by the disk read speed rather than network speed because the speed is almost the same on InfiniBand and GigE. 4) The block read speed is miserable compared to the normal continuous mode, so we must read the source audio file altogether before the encoding process. The bad performance is likely limited by disk performance because InfiniBand and GigE have the same performance.

Table IX shows the timing analysis of encoding one audio frame.

TABLE IX. TIME CONSUMPTION ENCODING ONE AUDIO FRAME

| Psychoacoustic model | MDCT | MS/LR Decision | Quantization Loop | Bitstream formatting | Total |
|---|---|---|---|---|---|
| 103us | 81us | 1us | 225us | 40us | 450us |
| 23% | 18% | 0 | 50% | 9% | 100% |

*2) Serial version (Method 1)*

In this experiment, the original LAME version 3.99.5 is compiled on the Iota cluster. The compiled executable encodes the source audio file. This experiment has been done three times, and an average time is recorded. The experiment results are shown in Table X.

TABLE X. EXPERIMENT RESULTS IN SERIAL VERSION

| Length | Original file size | MP3 file size | Compression ratio | Time used |
|---|---|---|---|---|
| 120 min | 1227 MB | 111 MB | 11:1 | 4:48 (248 s) |

*3) Segmented parallel version (Method 2)*

The source code is modified in this experiment to adapt to OpenMP, MPI, and UPC architectures. For MPI and UPC versions, the program will utilize 1 processor, 2 processors, 4 processors, 8 processors, and 16 processors. Since each node has 4 processors (or 4 cores), the number of nodes goes from 1 to 2 to 4. The experiment results for OpenMP are shown in Fig. 8 and Fig. 9.

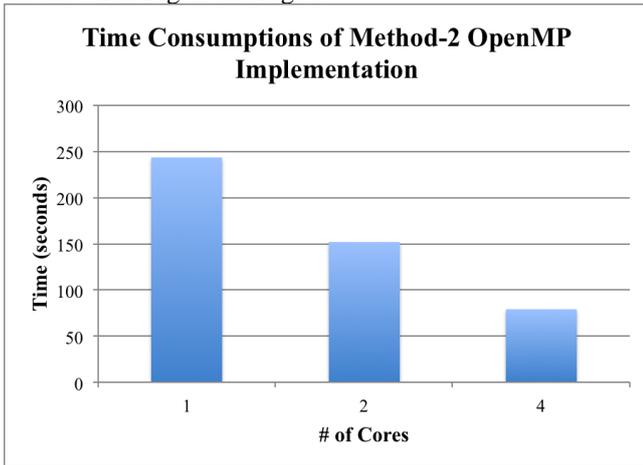

Fig. 8. Time consumptions in Method 2 (OpenMP)

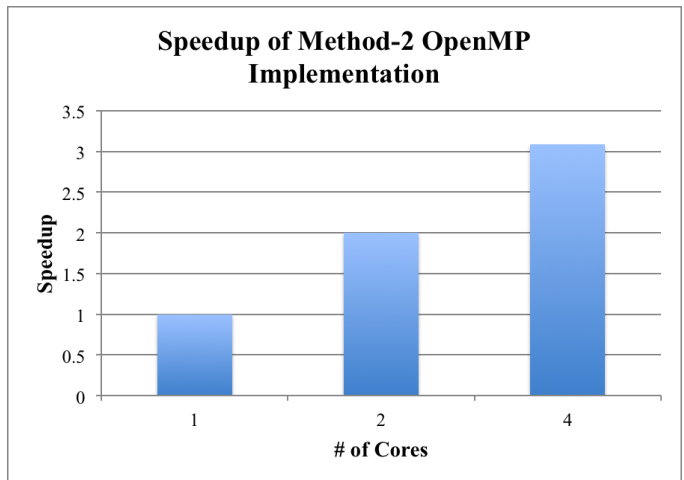

Fig.9. Speedups in Method 2 (OpenMP)

The experiment results for MPI and UPC architectures are shown in Fig. 10 and Fig. 11. For the MPI architecture, the application is set to run on two kinds of networks (Gigabit Ethernet and InfiniBand). For the UPC architecture, the application is only run on InfiniBand. Note that we don't use the round-robin scheme.

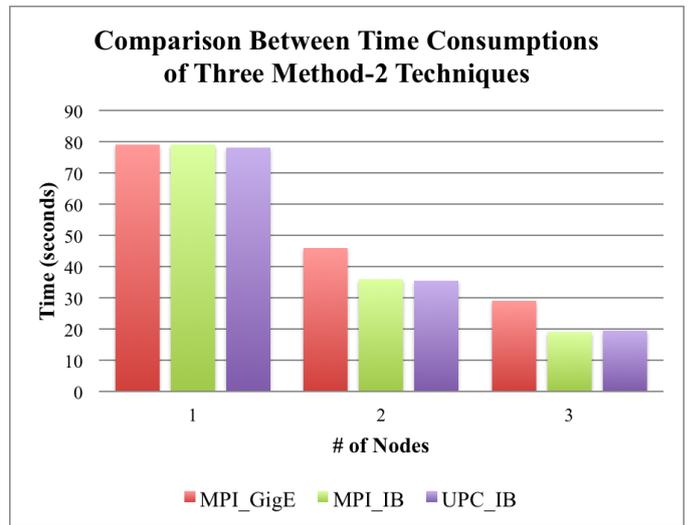

Fig. 10. Time consumptions in Method 2 (MPI and UPC)

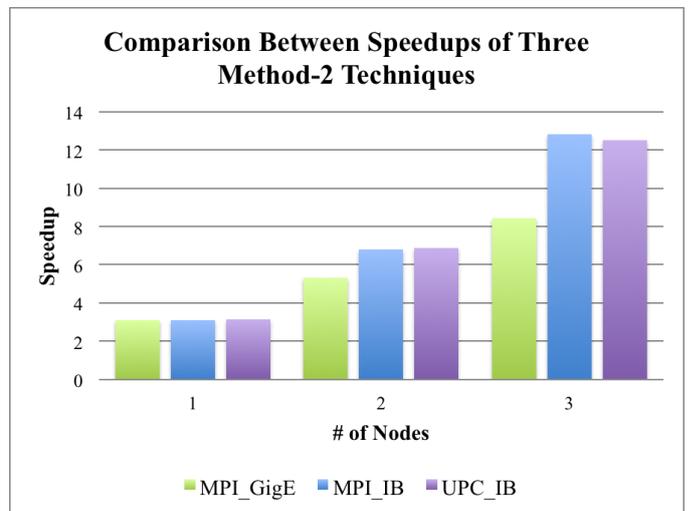

Fig. 11. Speedups in Method 2 (MPI and UPC)

*4) Work-sharing parallel version (Method 3)*

In this experiment, the source code is adapted according to Method 3. We used two architectures: MPI+OpenMP and UPC. The time consumption and corresponding speedup values are shown in Fig. 13 and Fig 14.

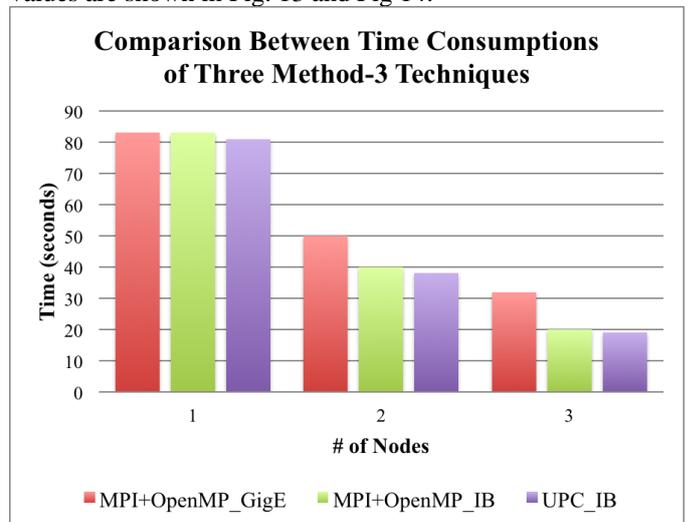

Fig. 13. Time consumptions in Method 3 (MPI+OpenMP Hybrid, UPC)

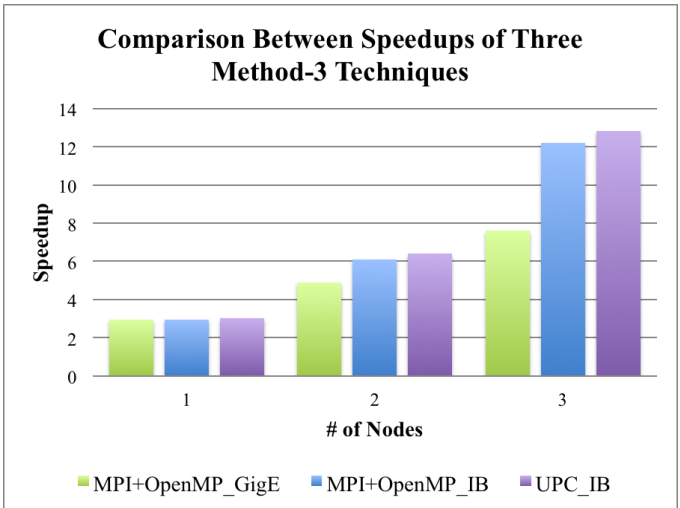

Fig. 14. Speedups in Method 3 (MPI+OpenMP Hybrid, UPC)

## V. ANALYSIS

### A. Huffman Coding

On Beta Cluster, the sequential algorithm had a total latency of 7.4105 seconds (TABLE I) and achieved a compression of 48.26 percent. The compression achieved was within the range that was expected. However, the latency was higher than the 5.1 to 5.5 seconds that was originally anticipated.

For the first parallel algorithm (METHOD 2), the latency on 1 node was about 5% higher than the sequential latency, which agrees with what was expected due to parallel overhead, while on 2 nodes, the latency was about 11 percent higher than the sequential and 6% higher than on 1 node. On 4 nodes, however, there was a 36% reduction in latency compared to the sequential. This was unexpected since, on 4 nodes, it was anticipated that the overhead in communication would be higher, thus increasing our latency. One possible explanation for this efficiency was the reduced datasets being worked on by each of 16 cores (4 nodes), which compressed its part of the dataset expeditiously and returned the results much shorter than would 8 or 4 cores. Of the three development libraries used, the algorithm using MPI+OpenMP gave a better performance on all three nodes. The compression achieved on each node was very similar to the sequential compression, with a <1% difference which was anticipated.

The second parallel algorithm (METHOD 3) achieved the best performance of the three methods implemented, with an amazing 83% reduction in latency compared to the sequential algorithm and compression of approximately 96% across all nodes. This performance was expected since METHOD 3 implemented a Burrows-Wheeler Transformation (BWT) on the dataset before applying Huffman Compression. The BWT transform reversed the text order in the dataset, creating runs of repeated characters, which allowed the Huffman algorithm to quickly and efficiently compress the data and return the results. The latencies across all nodes using this implementation were within one-tenth of a second among each other, with the smallest latency being achieved by the MPI Only application.

A very similar situation evolved when using the Iota cluster with a few key differences. The sequential algorithm took approximately 5.6 seconds to complete (most likely due to more powerful cores). This was much closer to the expected latency and a 32 percent improvement to that obtained using Beta Cluster. The compression achieved sequentially was 50.1 %, about 3% higher than that obtained with Beta.

The first parallel algorithm (METHOD 2) implemented on Iota had an approximately 8% parallel overhead compared to its serial counterpart, while on 2 nodes, the latency spiked to about 50% of its serial time. This behavior was similar to that in Beta when executing on 2 nodes. However, the latency was 17% less than the sequential time when executing on 4 nodes. This value was slightly better than that obtained for Beta, partly due to the InfiniBand implementation on Iota. The compression achieved across all three nodes was approximately 49%, consistent with the serial implementation, and the best-performing algorithm was the application of MPI+OpenMP.

The second parallel application (METHOD 3) had a 69% better latency performance than its serial counterpart when using 4 nodes and a parallel overhead of about 7 %. The compression achieved was approximately 96% due to applying the Burrows-Wheeler Transform as previously described for Beta. The MPI Only application achieved the best results for this method.

Fig. 5 shows a graphical representation of the latencies between the Beta and Iota Clusters. As can be seen from the chart, Iota had a much lower (better) sequential latency than Beta and also outperformed Beta when using 1 Node across the two parallel algorithms. When using 2 nodes, however, Beta slightly edged-out Iota, but the two clusters performed very similarly using 4 nodes. The two clusters obtained similar compression ratios for each node but it seems that implementations with MPI+OpenMP performs better on Beta, while UPC and MPI Only do better on Iota.

### B. LZSS

LZSS plays favorites. It performed, with respect to compression ratio, for JPEGs and random sequences but worked wonderfully for FPGA bitstreams. This is most likely because JPEGs have already been condensed, and, just as the name indicates, random sequences have very little repetition in them. With bitstreams, there are a lot of empty areas (all 0s) that could easily be compressed.

Also, since this method adds bits to the file, there were plenty of cases where the "compressed" file was larger than the original. This happens when there is repetition to exploit; the encode bits get added with no benefit.

Because of these encode bits, it is beneficial to use a reference even when the number of bits being replaced equals the sum of the offset and length bits because it prevents additional encode bits from possibly being added to the output file.

The multiple initial window values (also called dictionaries) are not normally stored in the file because that is a waste of space. The initial value is assumed to be known. Unfortunately, this only works if the perfect-sized, perfectly populated dictionary exists. The file itself is the best indicator of what kinds of patterns will appear later in the file. As shown with these experiments, although in the short run, storing this un-compressed information is a detriment, overall, it will increase the number of matches with the dictionary enough to over-compensate for that. The best results were always for that dictionary possibility.

There is a delicate balance between making the dictionary bigger (which increases the number of offset bits necessary to address it) and keeping it too small (where there is very little chance of having a match).

### C. MP3 Coding

The experiment results show that the MP3 encoding process can be parallelized. Moreover, the base for the parallelization is the data independence between two MP3 frames and two audio channels.

Audio compression is a data-intensive process. Thus the speed at which we move data around is crucial. From experiment 1) we can see that the performance of the disk limits the initial data read process. The disk may only be read at around 110 MB/s, less than Gigabit Ethernet's throughput. Moreover, the resulting MP3 file is small compared to the throughput. This is why in all situations, the performances of the Gigabit Ethernet and InfiniBand versions are the same when there is only 1 node (or 4 processors). Of course, as the number of nodes increases, those nodes have to send data to each other. Their performance will start to have a big difference.

Although the throughput is very different under Gigabit Ethernet and InfiniBand, the performance when reading a few bytes continuously is miserable. Reading a couple of bytes every cycle and continuously reading many cycles is not a problem on a single-node system. However, it has a great impact on parallel applications. If not fixed, this would slow the data reading as much as 50 times, which means the processors are effectively waiting for data all the time. So we have to read all the data continuously before we start compressing. Writing is less problematic because the resulting MP3 file is usually much smaller than the original.

In the OpenMP version of Method 2, we can see that the speedup is almost 2 when 2 cores are running; but it is a little over 3 when 4 cores are running. So the overhead is huge. The disk performance likely causes this: it simply cannot keep up with four simultaneously.

In the MPI and UPC versions of Method 2, we can see that the speedup of 2 nodes (8 processors) is more than 6, or twice that of OpenMP. This means the likely fault above is mitigated by moving the audio file to memory after reading. It can also be observed that the InfiniBand version performs better than the Gigabit Ethernet version, which aligns with our previous expectations.

In the MPI+OpenMP Hybrid and UPC versions of Method 3, it is observed that the UPC version performs better. This is caused by the UPC having better control over how the data is distributed. Moreover, compared with Method 2, the performance is about the same, with the UPC InfiniBand version better than the best performance in Method 2. This is not surprising because splitting the left and right channels and combining them has more memory implications. In UPC, we can set the memory of a two-dimensional array, which is exactly how dual channel data are stored, and bind to each thread exactly. This is why UPC performs better than MPI+OpenMP in this method. But it requires a lot more fine-grained coding and debugging.

## VI. CONCLUSIONS

### A. Huffman Coding

The analysis results done in parallelizing Huffman algorithms for text files indicate that significant compression latency improvements can be achieved by parallelizing the compression algorithms. Moreover, compression ratios greater than 90% are achievable by applying a Burrows-Wheeler Transform to the dataset before running the Huffman algorithm in parallel. Also, implementing MPI+OpenMP in parallel applications yields the most efficient results.

### B. LZSS

With an increase in cores, both parallel segmenting and CDC show an increase or no change in the compressed file size. Also, in terms of execution time, after 8 cores, the execution times level off, and by 32 cores, it has started increasing again. To maximize the benefit of time and space, running the algorithm in parallel on 8 processors would be best.

Surprisingly, with the number of tests done, LZSS Method 2 and Method 3 split the best compression ratios 50/50. It is clear, however, that Dictionary 2 saves much more space than it costs.

### C. MP3 Coding

For the MP3 coding part, two tasks have been done:

1) We parallelized the MP3 encoding process by feeding different parts of the source file to different processors. And then we concatenate them together.

2) We parallelized the MP3 encoding process by letting two worker processors work on two adjacent left channel frames while another two worker processors on one node work on two adjacent right channel frames.

The experiment results show that the MP3 encoding process can be parallelized. Moreover, the base for the parallelization is the data independence between two MP3 frames and two audio channels. The latter one is set manually so that we can speed up the encoding process. Since we did not break down the encoding process too small, the impact on the compression ratio is minimal.

Of all three methods, the UPC version of Method 3, using InfiniBand, performs best. However, it is worth noting that it requires more work than the MPI+OpenMP Hybrid version. This is because we can better tune the memory locations and memory access patterns of pointers in UPC. This requires more code and more complex debugging. The performance gained, however, is not significant. So MPI+OpenMP Hybrid version of Method 3 may be a better choice regarding effort and output.

In the future, we may try another approach: do parallelization at a finer level. For example, the psychological modeling and quantization loop are the two biggest time-consuming steps in the process. [11] We may try to parallelize them directly or indirectly. We may choose to parallelize the quantization loop or try to parallel the FFT used in the psychological modeling. Alternatively, we may even try to do global Huffman encoding. The goal is the speedup.